\documentclass[twocolumn,showpacs,superscriptaddress,prb]{revtex4}
\usepackage{psfig}
\usepackage{graphicx}
\usepackage{amsmath}
\usepackage{amssymb}
\usepackage{amsfonts}
\usepackage{bm}

\newcommand{\eqname}[1]{\label{eq:#1}}
\newcommand{\bgar}{\begin{eqnarray}}
\newcommand{\enar}[1]{\label{eq:#1}\end{eqnarray}}

\newcommand{\kk}{ {\bf k}}

\newcommand{\xx}{ {\bf x}}

\newcommand{\yy}{ {\bf y}}

\newcommand{\eq}[1]{(\ref{eq:#1})}
\newcommand{\al}[1]{^{(#1)}}
\newcommand{\Psihd}{\hat\Psi^\dagger}

\newcommand{\Psih}{\hat\Psi}

\newcommand{\Hamilt}{{\mathcal H}}

\begin{document}

\title{Spontaneous microcavity-polariton coherence across the
  parametric threshold: Quantum Monte Carlo studies}

%\affiliation{BEC-INFM, Universit\`a di Trento, 38050 Povo, Italy}
%\affiliation{Laboratoire Kastler Brossel, \'Ecole Normale
%Sup\'erieure, 24 rue Lhomond, 75005 Paris, France}

\author{Iacopo Carusotto}
\affiliation{BEC-INFM and Dipartimento di Fisica, Universit\`a di Trento,
  38050 Povo, Italy} 
\email{carusott@science.unitn.it}

\author{Cristiano Ciuti}
 \affiliation{Laboratoire Pierre Aigrain, \'Ecole Normale Sup\'erieure, 24 rue
 Lhomond, F-75231 Paris Cedex 05, France}

\begin{abstract}
We investigate the appearance of spontaneous coherence in the
parametric emission from planar semiconductor microcavities in the
strong coupling regime.  
Calculations are performed by means of a Quantum Monte Carlo technique
based on the Wigner representation of the coupled exciton and
cavity-photon fields. 
The numerical results are interpreted in terms of a non-equilibrium phase
transition occurring at the parametric oscillation threshold:
below the threshold, the signal emission is incoherent, and both the
first and the second-order coherence functions have a finite
correlation length which becomes macroscopic as the threshold is approached.
Above the threshold, the emission is instead phase-coherent over the
whole two-dimensional sample and intensity fluctuations are suppressed.
Similar calculations for quasi-one-dimensional microcavities show that
in this case the phase-coherence of the signal emission has a finite
extension even above the threshold, while intensity fluctuations are
suppressed.
\end{abstract}

%\vspace{1cm}1


\date{\today}

\maketitle

\section{Introduction}
\label{sec:Intro}

The recent progresses in the growth and manipulation techniques of
semiconductor heterostructures have recently led to the realization of
planar semiconductor microcavities in which a cavity-photonic mode is
strongly coupled to an excitonic transition, so that the resulting
elementary excitations consist of linear superpositions of
cavity-photon and exciton, the so-called {\em
  polaritons}~\cite{MC_Review,MC_Review2,MC_ReviewSSC}.

The polaritons satisfy the Bose statistics, at least in a low-density
regime, and  have recently started to be considered as potential
candidates for the study of many-body effects,
e.g. superfluidity~\cite{OurSuperfluidity}, in regimes completely
different from the ones usually considered in experiments with Helium
fluids~\cite{ManyBody} or ultracold atomic 
gases~\cite{AtomicBEC}.

A further advantage of polaritonic systems consists of the fact that
polaritons can be optically injected in the cavity by simply shining an
incident light beam onto it, and their properties can then be inferred
from the observation of the emitted light. In particular, the coherence
properties of the light emission closely reproduce the ones of the
in-cavity polaritonic field~\cite{AugustinCoh}.

Thanks to their excitonic component, polaritons have strong binary
interactions, which have been exploited to obtain spectacular
polariton amplification and optical parametric oscillation
effects~\cite{MC_ReviewSSC,ParametricEmission,Bragg,Yama_g2}.
Differently from conventional optical resonators with a discrete set of
optical modes, planar microcavities are spatially extended systems
with a continuum of transverse modes and the parametric oscillation
process automatically selects the most favourable one in which to
generate the coherent signal field. 

Simultaneously with these experiments, an active debate has taken
place in the community about the relation between the appearance of
spontaneous coherence in optical parametric oscillation in planar
microcavities and Bose-Einstein condensation of
excitons~\cite{Excit_BEC}.
The present paper is devoted to shine light on analogies and
differences between the two phenomena: both
of them are in fact characterized by the appearance of coherence and
long-range order in a Bose field~\cite{AtomicBEC,CohBose,PenroseOnsager}.
There is however a major difference: while Bose-Einstein condensation is
an equilibrium phenomenon, parametric oscillation takes place in
intrinsically out-of-equilibrium systems, where the stationary state 
arises from a dynamical equilibrium between the driving by the
pump and the dissipative effects due to losses.
From this point of view, the parametric oscillation threshold is closely
related to the laser threshold~\cite{OptPhaseTrans,ScullyZubairy},
although standard laser cavities have a finite number of discrete modes.

In order to perform quantitative calculations, a Monte Carlo
technique based on the Wigner representation of a quantum
field~\cite{WallsMilburn} has 
been generalized to the  case of a pair of coupled excitonic and cavity-photonic
fields and the resulting numerical code has been used to 
calculate the one-time correlation functions of the
in-cavity field in the stationary state, in particular the first- and
second-order ones.
As already mentioned, these univocally determine the spatial coherence
properties of the emitted light which can be measured in the experiments.

The advantages of the Wigner-Monte Carlo approach with respect to
previous works on the subject~\cite{Ciuti_Review} are manyfold: first of all, we keep
track in a complete way of the 
fluctuations of the polaritonic field, even in the critical region
around the threshold, where fluctuations are large and the linearized
approaches around the mean-field~\cite{CiutiParamLum,Bragg,Savona} fail.
Furthermore, no few-mode
approximation~\cite{Savona,CiutiOPO,WhittakerOld,Whittaker} is here
performed: the parametric emission into all available modes is automatically taken
into account, as well as the correlations between all modes.

Above threshold, the signal emission in planar microcavity geometries
occurs in the single, most favourable, mode (which the numerical code
is able to automatically select) and therefore shows the typical spatial 
coherence properties of a Bose-Einstein condensate. As it is not
inherited from the pump beam, this coherence is a {\em spontaneous} one.

On the other hand, the parametric emission from reduced-dimensionality
systems such as photon wires~\cite{PhotonWires} shows dramatically
different properties: as long-wavelength fluctuations of the signal
phase are here able to destroy the long-range spatial coherence, the
threshold is no longer well defined and is replaced by a smooth
crossover where a local order sets in and the intensity fluctuations
of the signal emission are suppressed. 
Such a behaviour is closely related to the so-called
quasi-condensation phenomenon of equilibrium Bose gases in reduced
dimensionalities~\cite{LowD,quasicond,quasicond2}.

Related questions about the spontaneous appearance of long-range
coherence and the effect of a reduced dimensionality have been
recently adressed~\cite{QuantumImagesOPO,ZambriniPRA} for the
simpler case of a parametric oscillator formed by a planar resonator
containing a slab of generic $\chi\al{2}$ medium without any excitonic
resonance.

The plan of the present paper is the following: in sec.\ref{sec:Hamilt} we
introduce the field-theoretical Hamiltonian of the system and we
relate the properties of the emitted light to the ones of the
in-cavity polaritonic field.
In sec.\ref{sec:MFT}, we work out the corresponding mean-field theory
and we use it to identify the parameter range which is most suited
for the following analysis.
In sec.\ref{sec:Wigner}, we review the Wigner representation formalism
and we generalize it to the case of coupled cavity-photon and exciton
fields. We also describe the principles of the numerical code used to
obtain the results which are discussed in the following.
In sec.\ref{sec:Coherence2D}, the central numerical results of the
paper are shown: the behaviour of the first- and second-order
coherence properties of the signal emission are presented for pump
parameters spanning across the parametric threshold.
These results are the basis of the discussion of the following
sec.\ref{sec:AnalogyBEC}, where analogies and differences between our
driven-dissipative system and the Bose-Einstein condensation phenomenon of
equilibrium statistical mechanics are pointed out.
Sec.\ref{sec:Modulational}, an alternative interpretation of the
parametric threshold in terms of modulation instability and
spontaneous breaking of the translational symmetry is given and its
consequences on the intensity correlation function are put
forward.
The effect of a reduced dimensionality is discussed in
sec.\ref{sec:1D}, where the link with the concept of quasi-condensate
is evidentiated.
Finally, in sec.\ref{sec:Conclu} we give our conclusions, and we
sketch the perspectives of the work.

\section{The system Hamiltonian}
\label{sec:Hamilt}

A model which is commonly used to describe a planar microcavity
with a quantum well excitonic resonance strongly coupled to the cavity
mode is based on the field-theoretical Hamiltonian 
\cite{Ciuti_Review}:
\begin{multline}
  \label{eq:Hamilt_tot}
  \Hamilt=\int\!d\xx\,\sum_{ij=\{X,C\}}
\Psihd_{i}(\xx) \,\Big[{\mathbf
  h}^{0}_{ij}+V_{i}(\xx)\,\delta_{ij} \Big]\,\Psih_{j}(\xx)\\
%+\sum_{i} V_{i}(\xx)\,\Psihd_{i}(\xx)\,\Psih_{i}(\xx) \\
+\frac{\hbar g}{2}\int\! d\xx\,\Psihd_{X}(\xx)\,\Psihd_{X}(\xx)\,
\Psih_{X}(\xx)\,\Psih_{X}(\xx)+\\
+\int\!d\xx\,\hbar\beta_{inc}\,E_{p}(\xx,t)%\,e^{i(\kk_{p} \xx-\omega_{p}t)}
\,\Psihd_{C}(\xx)+\textrm{h.c.}~,
\end{multline}
where $\xx$ is the in-plane spatial position.
The field operators $\Psi_{X,C}(\xx)$ respectively describe excitons
($X$) and cavity photons ($C$) and, provided the exciton density is
much lower than the saturation density $n_{sat}$~\cite{Ciuti_Review},
they satisfy Bose commutation rules,  
$[\Psih_i(\xx),\Psihd_j(\xx')]=\delta^2(\xx-\xx')\,\delta_{ij}$.
For the sake of simplicity, we consider here a single polarization
state, but a generalization of \eq{Hamilt_tot} to include several
polarization states and describe the spin dynamics discussed
e.g. in Ref.\onlinecite{Spin} is straightforward. 

The linear Hamiltonian ${\mathbf h}^0$ is:
\begin{equation}
  \label{eq:Hamilt_lin}
  {\mathbf h}^0=
\hbar \left(
\begin{array}{cc}
\omega_{X}(-i\nabla) & \Omega_R \\
\Omega_R & \omega_C(-i\nabla)
\end{array}
\right),
\end{equation}
where $\omega_{C}(\kk)=\omega_{C}^0\,\sqrt{1+{\kk^2}/{k_z^2}}$ is
the cavity dispersion as a function of the in-plane wavevector
$\kk$ and $k_z$ is the quantized photon wavevector in the growth
direction. 
The exciton dispersion is instead much weaker and can be well
approximated by a constant $\omega_X(\kk)=\omega_X$.
The quantity $\Omega_R$ is the Rabi frequency of the exciton-cavity
photon coupling: the eigenmodes of the linear Hamiltonian
\eq{Hamilt_lin}  are therefore linear combination of the cavity
photonic and excitonic modes. They are called upper ($UP$) and lower
($LP$) polariton and their dispersion is  given by
$\omega_{LP(UP)}(\kk)$. 

The nonlinear interaction term is due to exciton-exciton collisional
interactions and can be modelled by a repulsive ($g>0$) contact
potential. The anharmonic exciton-photon coupling has a negligible
effect in the regime considered in the present study~\cite{Ciuti_Review}.
$V_{X,C}(\xx)$ are external potential terms acting on the excitonic
and photonic fields and can be used to model etched
cavities~\cite{PhotonWires}.

Differently from other Bose systems such as ultracold trapped atoms
or liquid Helium, our polaritonic system is intrinsically out of
equilibrium. 
Polaritons are in fact injected in the cavity by incident light
beams and decay through several dissipative channels.
In the present paper, polaritons are coherently injected in the cavity 
by a coherent {\em pump} laser, which is taken into account in the
Hamiltonian \eq{Hamilt_tot} by the driving term proportional to the
incident electric field $E_{p}(\xx,t)$ of the pump.
For a given cavity, the coefficient $\beta_{inc}$ can be calculated
from the reflectivity of the front mirror through which the cavity is
excited~\cite{WallsMilburn}.
Both the spatial and the temporal dependences of $E_p(\xx,t)$ can be
experimentally controlled, so to obtain a variety of different temporal
(e.g. monochromatic, bichromatic, pulsed, etc.) and spatial
(e.g. plane wave, gaussian, optical vortex, etc.)
patterns~\cite{IncidentPatterns}.

Dissipative effects lead to a damping of both the excitonic and cavity
photonic fields. Using the standard quantum theory of damping, they
can be modelled by additional terms in the master equation giving the
time-evolution of the density matrix $\rho$ of the
system~\cite{WallsMilburn}: 
\begin{multline}
\eqname{damping_rho}
\frac{d\rho}{dt}=-\frac{i}{\hbar}\big[\Hamilt,\rho\big]+
\int\!\frac{d\kk}{2\pi}\,\sum_{i=\{X,C\}}
\gamma_i(\kk)\,\Big[\Psih_i(\kk)\,\rho\,\Psihd_i(\kk)\\
-\frac{1}{2}\big(\Psihd_i(\kk)\,\Psih_i(\kk)\,\rho+\rho\,\Psihd_i(\kk)\,\Psih_i(\kk)\big)
\Big].
\end{multline}
The operators $\Psih_{X,C}(\kk)$ are the Fourier transform of the
field operators $\Psih_{X,C}(\xx)$.
In the present paper, we shall consider the case of
momentum-independent $\gamma_{C,X}$ which summarize dissipative
effects of different origin, e.g. excitonic scattering and
non-radiative recombination, background absorption by the
material forming the structure as well as radiative emission into the
continuum of modes outside the cavity.

From the experimental point of view, information on the state of the
polariton field in the cavity is generally retrieved by detecting and
characterizing the light emission from the cavity through either the
front or the back mirror.
In particular, the normally-ordered expectation values for the light
emitted through the back mirror (called in the following
{\em transmitted} light)
\footnote{The analysis of the reflected light requires more attention:
the field emitted by the cavity through the front mirror has in fact to be
added to the direct reflection of the pump beam on the front mirror.}
are proportional to the corresponding ones for the photonic component
  of the in-cavity polaritonic field.
For instance, the near-field amplitude of the transmitted field is 
\begin{equation}
\big\langle {\hat E}_{tr}^\dagger(\xx)\,{\hat
  E}_{tr}(\xx')\big\rangle=|\beta_{tr}|^2\, 
\big\langle \Psihd_C(\xx)\,\Psih_C(\xx') \big\rangle,
\end{equation}
where the coefficient $\beta_{tr}$ relating the transmitted field to
the in-cavity one has to be calculated from the transmittivity of the
back mirror~\cite{WallsMilburn}.
Similar formulas hold in the far-field. The far-field angular pattern
is proportional to the in-cavity occupation number: 
\begin{multline}
\big\langle
{\hat E}_{tr}^\dagger(\kk)\,{\hat E}_{tr}(\kk)
\big\rangle=|\beta_{tr}|^2\,n_C(\kk)=\\
=|\beta_{tr}|^2\,\big\langle \Psihd_C(\kk)\,\Psih_C(\kk) \big\rangle
\end{multline}
and analogous expressions hold for its higher-order, normally-ordered
moments such as:
\begin{multline}
\big\langle
{\hat E}_{tr}^\dagger(\kk)\,{\hat E}_{tr}^\dagger(\kk')\,{\hat E}_{tr}(\kk')\,
{\hat E}_{tr}(\kk) \big\rangle=\\
=|\beta_{tr}|^4\,
\big\langle \Psihd_C(\kk)\,\Psihd_C(\kk')\,\Psih_C(\kk')\,\Psih_C(\kk)
\big\rangle.
\end{multline}
This procedure has recently been used to experimentally determine the
first- and second-order correlation functions of the polaritonic field
by measuring the coherence and noise properties of the emitted
light~\cite{AugustinCoh}.

\section{Mean-field theory}
\label{sec:MFT}

A simple way of coping with the field Hamiltonian \eq{Hamilt_tot} is
to study the corresponding classical field theory once the 
quantum field operators $\Psi_{X,C}(\xx)$ are replaced with classical 
${\mathbf C}$-number fields $\psi_{X,C}(\xx)$ defined as the
expectation value of the quantum field $\psi_{X,C}(\xx)=\big\langle
\Psi_{X,C}(\xx)\,\big\rangle$.
For a system of weakly-interacting bosons like microcavity polaritons,
fluctuations are generally weak provided the mean-field solution is
dynamically stable, so that the mean-field solution provides accurate
predictions for the coherent part of the polariton field, and
therefore the coherent emission from the microcavity.
This includes e.g. resonant Rayleigh scattering in the presence of
defects~\cite{OurSuperfluidity}. 
In the present paper, the mean-field prediction for the threshold
intensity will be used as a guideline in the choice of parameters to
be used in the numerical calculations. 

As we are most interested in the basic properties of
the parametric instability threshold, we have tried to identify a
configuration where the intrinsic phenomenology is not masked by
additional effects. 
In particular, we have considered the case of a homogeneous cavity
with no external potential $V_{X,C}=0$ and a transversally
homogeneous, monochromatic and cw laser field of frequency
$\omega_{p}$ which incides on the cavity at an angle 
$\theta_p$ with the normal to the cavity plane.
Provided the beam actually used in the experiment is wide enough in
the transverse direction, this approximation is expected to be
accurate.
The pump electric field has then the simple form:
\begin{equation}
E_p(\xx,t)=E_p\,e^{i(\kk_p\cdot\xx-\omega_p t)}.
\end{equation}
the wavevector $\kk_p$ being fixed by the incidence angle through
$k_{p}=\sin\theta_p\;\omega_{p}/c$. The incident intensity $I_p$ is
proportional to $|E_p|^2$:
\begin{equation}
I_p=\frac{c}{2\pi}\,|E_p|^2.
\end{equation}

Given the translational symmetry of the system, we can look for stationary
solutions in the same plane-wave form 
$\psi_{X,C}(\xx,t)=\psi_{X,C}^{ss}\,e^{i(\kk_p\xx-\omega_p t)}$ as the
incident field. The coefficients $\psi_{X,C}^{ss}$ can be obtained from
the equation of state:
\begin{eqnarray}
\Big(\omega_{X}(\kk_{p})-\omega_{p}-\frac{i}{2}\gamma_X+
g\,|\psi_X^{ss}|^2\Big)\,
\psi^{ss}_{X}+\Omega_R\,\psi_C^{ss}=0  \label{eq:ss_X} \\
\Big(\omega_{C}(\kk_{p})-\omega_{p}-\frac{i}{2}\gamma_C \Big)\,
\psi^{ss}_{C}+\Omega_R\,\psi_X^{ss}=-\beta_{inc} E_p  \label{eq:ss_C}
\end{eqnarray}
and the stability of the solution has to be verified by looking at the
eigenvalues of the Bogoliubov matrix~\cite{OurSuperfluidity}:

\begin{widetext}
\begin{equation}
  \label{eq:Bogo_L}
  {\mathcal L}=
\left(
\begin{array}{cccc}
\omega_X+2g\,|\psi_X^{ss}|^2-  \omega_p -\frac{i\gamma_X}{2} &
\Omega_R & g\,\psi^{ss\,2}_X\,e^{2 i \kk_p\xx} & 0 \\
\Omega_R & \omega_C(-i\nabla)-  \omega_p - \frac{i\gamma_C}{2} & 0 & 0 \\
-g\,\psi^{ss\,*\,2}_{X}\,e^{-2 i \kk_p\xx} & 0 &
-\big(\omega_X+2g\,|\psi_X^{ss}|^2\big) +
\omega_p-\frac{i\gamma_X}{2} &
-\Omega_R \\
0 & 0 & -\Omega_R & -\omega_C(-i\nabla)+ \omega_p-\frac{i\gamma_C}{2}
\end{array}
\right).
\end{equation}
\end{widetext}
If the imaginary parts of all the eigenvalues
$\omega_{UP,LP}^\pm(\kk)$ of the Bogoliubov matrix ${\mathcal L}$ are
negative for all $\kk$, the mean-field solution is 
dynamically stable; if at least one of them is positive, a different,
spatially structured, stationary solution has to be found, 
e.g. with a sinusoidal intensity profile, and its stability has again to
be verified by means of the appropriate Bogoliubov matrix.
An analytical study of the mean-field stationary state in the
parametric oscillation regime when translational symmetry is broken is
beyond the scope of the present work.
Simple discussions under a simple three-mode approximation taking into
account only the pump, the signal and the idler modes can be found e.g. 
in Ref.\onlinecite{Savona,CiutiOPO,WhittakerOld,Whittaker}.

\begin{figure}[htbp]
\includegraphics[width=3.3in,clip]{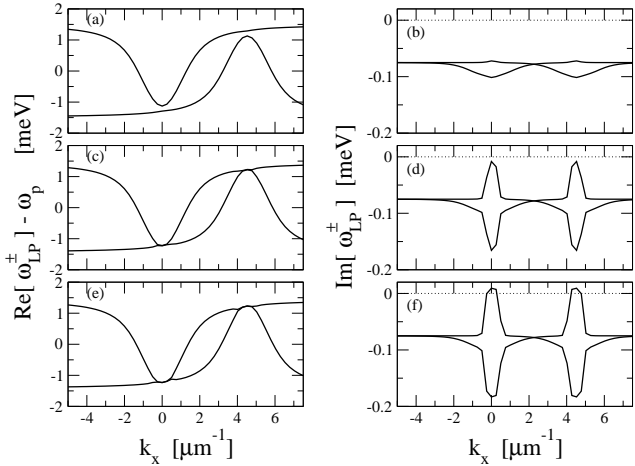}
\caption{Real (left) and imaginary (right) parts of the Bogoliubov
  dispersion for different values of the pump frequency  
%$\omega_p=1.3987\,\textrm{eV}$  
$\hbar\delta_p=\hbar(\omega_p-\omega_{LP}(\kk_p)\,)= -0.65 \,\textrm{meV}$  
(a,b), 
%$1.39881\,\textrm{eV}$ 
$-0.54\,\textrm{meV}$ 
(c,d),
%  $1.39886\,\textrm{eV}$ 
$-0.49\,\textrm{meV}$ 
(e,f). 
At the mean-field threshold
%$\omega_p\al{c}\approx 1.39883$. 
$\hbar\delta_p=\hbar\delta_p\al{c}\approx -0.52\,\textrm{meV}$,
the mean-field energy  $\hbar g\,|\psi_{X}|^2\simeq
0.13\,\textrm{meV}$.
For the sake of clarity, only the branches corresponding to the LP
  have been traced, the UP ones being far away 
  in energy. Cavity parameters:
  $\hbar \omega_C^0=\hbar \omega_X=1.4\,\textrm{eV}$, 
$2\,\hbar \Omega_R=5\,\textrm{meV}$,
  $\hbar \gamma_{X}= 0.15,\,\textrm{meV}$, 
$\hbar\gamma_C=0.24\,\textrm{meV}$.
The lower polariton energy at linear regime is
$\hbar\omega_{LP}(\kk_p)=1.39935\,\textrm{eV}$.
\label{fig:Bogo}}
\end{figure}

In order to keep the physics the simplest, we have chosen a fixed
value of the pump intensity $I_p$ 
and we have spanned its frequency 
$\omega_p$ from far below the lower polariton frequency at linear
regime $\omega_{LP}(\kk_p)$ in the upward direction.
The incidence angle $\theta_p$ has been taken slightly larger than the
so-called  magic angle, so that
$2\omega_{LP}(\kk_p)>\omega_{L}(0)+\omega_{LP}(2\kk_p)$. 
The corresponding Bogoliubov dispersions are plotted in
fig.\ref{fig:Bogo}: as expected, the first feature that takes place is
the parametric instability when the $\pm$ branches, which are coupled
by the anti-hermitian terms of \eq{Bogo_L}, touch around $\kk\approx
0$. This condition $2\omega_p=\omega_{LP}^+(0)+\omega_{LP}^+(2\kk_p)$
can be interpreted as the parametric process 
$(\kk_p,\kk_p)\rightarrow (0, 2\kk_p)$ becoming resonant on the final
state.

For our specific choice of $\theta_p$ and $\omega_p$, notice that this
happens at a value $\omega_p=\omega_p\al{c}$ still red-detuned with
respect to $\omega_{LP}(\kk_p)$.
This guarantees that we are still in an optical limiter regime and no
single-mode optical bistability effects~\cite{OurSuperfluidity} has
yet taken place.
In the following of the paper, the frequency $\omega_p\al{c}$ will be
designed as the parametric instability threshold and the wording
below/above threshold will refer to $\omega_p$ respectively red- or
blue-detuned with respect to $\omega_p\al{c}$. This definition of the
threshold in terms of frequency is alternative but equivalent to the
standard one in terms of the intensity and allows us to avoid those
complicate additional effects that have been pointed out
in Ref.\onlinecite{Whittaker}.
For notational simplicity, we shall also introduce the quantity
$\delta_p=\omega_p-\omega_{LP}(\kk_p)$; at the threshold
$\delta_p=\delta_p\al{c}<0$.

\section{The Wigner Monte Carlo method}
\label{sec:Wigner}

A more complete analysis of the parametric threshold can be carried out
by numerically solving the stochastic equations of the Wigner
representation of the quantum Bose field.
This technique allows one to study the dynamics of the field taking
fully into account the multi-mode structure of the field and
eventually the spontaneous breaking of the translational symmetry, as
well as the fluctuations around the mean-field solution, however large
they happen to be.
In the present section, we shall review the formalism of the Wigner
representation of the quantum Bose field as applied to a pair of
coupled photonic and excitonic fields with a quartic interaction term.
Recently, a similar Wigner approach has been applied to study the
dynamical properties of ultracold atomic gases~\cite{WignerAtoms}. 

The Wigner representation of the two-component quantum Bose field
$\Psih_i(\xx)$ ($i=\{X,C\}$) consists in a quasi-probability
distribution function $W[\psi_i(\xx)]$ in the functional space of the
${\mathbf C}$-number fields $\psi_i(\xx)$~\cite{WallsMilburn}.
Any numerical implementation requires one to work on a discrete and
finite spatial grid: in our simulations, a rectangular grid of ${\mathcal
  N}_x\times{\mathcal N}_y$ points with uniform grid spacings $\ell_{x,y}$
along respectively the $x,y$ directions has been used, with periodic
boundary conditions. 
The integration box has therefore sides $L_{x,y}={\mathcal
  N}_{x,y}\ell_{x,y}$.

Under a diluteness condition that can be written as $\gamma_i\gg
g/dV$ ($dV=\ell_x\ell_y$ being the cell volume), the evolution of the
quasi-probability distribution $W$ can be approximated in the
so-called {\em truncated Wigner
  approximation}~\cite{WignerAtoms,YakWigner} by a true 
Fokker-Planck equation with a positive-definite 
diffusion matrix and without higher-order derivative
terms, which guarantees the positivity of $W$ at all
times and the possibility of mapping the problem onto stochastic
differential equations for the fields $\psi_i(\xx)$:
\begin{widetext}
\begin{multline}
  \label{eq:GPE}
  d
\left(
  \begin{array}{c}
\psi_{X}(\xx) \\ \psi_{C}(\xx)
  \end{array}
\right)=\\
=-\frac{i}{\hbar}\,\left\{
\left(
  \begin{array}{c}
0 \\ \hbar\beta_{inc}\,E_{p}(\xx,t)
  \end{array}
\right)+\left[
{\mathbf h}^0+
\left(
\begin{array}{cc}
V_X(\xx)+\hbar g
\Big(|\psi_X(\xx)|^2-\frac{1}{dV}\Big)-\frac{i\gamma_X}{2} & 0 
\\ 0 & V_C(\xx)-\frac{i\gamma_C}{2}
  \end{array}
\right)
\right]
\left(
  \begin{array}{c}
\psi_{X}(\xx) \\ \psi_{C}(\xx)
  \end{array}
\right)\right\}\,dt+\\
+
\frac{1}{\sqrt{4\,\Delta V}}
\left(
  \begin{array}{c}
\sqrt{\gamma_X}\,dW_X(\xx) \\ \sqrt{\gamma_C}\,dW_C(\xx)
  \end{array}
\right).
\end{multline}
\end{widetext}
where $dW_{X,C}(\xx)$,  are zero-mean, independent, complex-valued,
white noise terms such that 
\begin{eqnarray}
\overline{dW_i(\xx)\,dW_j(\xx')}&=&0 \\
\overline{dW_i(\xx)\,dW^*_j(\xx')}&=&2\,dt\,\delta_{\xx,\xx'}\,\delta_{ij}.
\end{eqnarray}
In the most relevant case of an initially empty microcavity, the
initial fields at $t=0$ are independent zero-mean complex Gaussian
variables such that:
 \begin{eqnarray}
\overline{\psi_i(\xx)\,\psi_j(\xx)}&=&0 \\
\overline{\psi_i(\xx)\,\psi^*_j(\xx)}&=&\frac{1}{2\,\Delta V}\,\delta_{ij}
\end{eqnarray}

In a practical Wigner-Monte Carlo simulation, the Wigner distribution of the
initial state at $t=0$ is sampled with a large number ${\mathcal N}$
of independent realizations (the data shown in the present paper have
been calculated from a number ${\mathcal N}$ of realizations of the
order of a few thousands). They are then let evolve according to the
stochastic equation \eq{GPE} with a random noise which is obviously
different for each realization.
 
The expectation value of any observables at time $t$ is obtained as an
average over the different realizations of the corresponding classical
quantity, paying attention to the fact that the moments of the Wigner
function give the expectation value of totally symmetrized operators, e.g.:
\begin{equation}
\big\langle O_1\ldots O_N \big\rangle_W=
%\frac{1}{N!}\sum_P \textrm{Tr}[{\hat
%    O}_{P(1)}\ldots {\hat O}_{P(N)}\,\rho]=\\
\frac{1}{N!}\sum_P\big\langle {\hat O}_{P(1)}\ldots{\hat O}_{P(N)}
    \big\rangle, 
\end{equation}
where the sum is taken over all the permutations $P$ of $N$
objects, each of the $O_\alpha$'s stands for some field component
$\psi_i(\xx)$ or $\psi_i^*(\xx)$ and ${\hat O}_\alpha$ is the
corresponding field  
operator. As a simplest example, 
\begin{equation}
  \label{eq:N_k}
  \big\langle \psi_C^*(\xx)\,\psi_C(\xx')
  \big\rangle_W=
\frac{1}{2}\,\big\langle\Psihd_C(\xx)\Psih_C(\xx')+\Psih_C(\xx')\Psihd_C(\xx)  \big\rangle
%=\big\langle\Psihd_\kk\Psih_\kk\big\rangle+\frac{1}{2}
\end{equation}
In the following of the paper, we shall use this Wigner-Monte Carlo
method to numerically study the one-time correlation functions of the
in-cavity fields in the stationary state. 
These are obtained by letting the evolution to run for a time interval
longer than all the characteristic time scales of the problem, until
the stationary state is attained (in practice an evolution time of at
most $5\,\textrm{ns}$ is used). At this point, the expectation
values of the observables are evaluated from the corresponding
averages over the Wigner distribution.
Note that we have chosen the excitation parameters in such a way that
the pump has a simple optical limiter behaviour and complex
bistability effects are avoided.

For instance, the first-order correlation function in the near-field
has the form:  
\begin{equation} 
\big\langle \Psihd_C(\xx)\,\Psih_C(\xx') \big\rangle=\Big[
\big\langle \psi_C^*(\xx)\,\psi_C(\xx')
  \big\rangle_W-\frac{\delta_{\xx,\xx'}}{2\,\Delta V}\Big],
\eqname{g1_wign}
\end{equation}
while the second-order one reads:
\begin{multline}
\big\langle \Psihd_C(\xx)\,\Psihd_C(\xx')\,\Psih_C(\xx')\,\Psih_C(\xx)
\big\rangle= \\ 
=\Big\langle
|\psi_C(\xx)|^2\,|\psi_C(\xx')|^2-\frac{1}{2\,\Delta 
  V}\big(1+\delta_{\xx,\xx'}\big) \\
\cdot\Big(\,|\psi_C(\xx)|^2+|\psi_C(\xx')|^2 -\frac{1}{2\,\Delta
  V}\Big)
  \Big\rangle_W.
\eqname{g2_wign}
\end{multline}
Similar formulas hold for the $\kk$-space observables which describe
the emission pattern in the far-field.

The study of multi-time correlation functions which are involved in
the temporal coherence properties requires the calculation of
non-trivial different-time commutators or a specific treatment of
quantum noise incident on the cavity~\cite{ZambriniPRA}. 
All these issues go beyond the scope of the present work and will be
the subject of a forthcoming publication.

\section{Coherence properties of the parametric emission}
\label{sec:Coherence2D}

The Wigner-Monte Carlo method reviewed in the previous section can be
applied to study the parametric emission for $\omega_p$ spanning
across the threshold. 
Depending on whether $\omega_p<\omega_p\al{c}$ or
$\omega_p>\omega_p\al{c}$, the emission shows in fact dramatically
different features which appear clearly in the angular intensity
pattern of the far-field emission as well as in its first- and
second-order coherence properties.

\subsection{Below the parametric instability threshold}

\subsubsection{Far-field angular pattern}

\begin{figure}[htbp]
\includegraphics[width=3.3in]{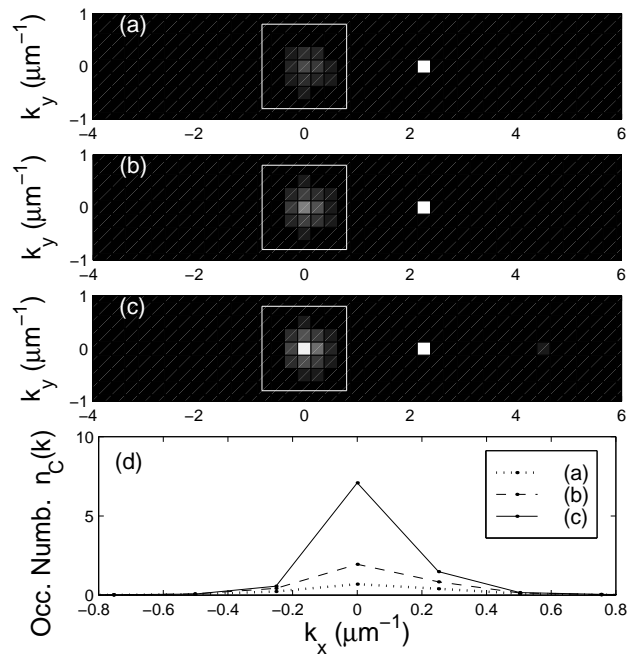}
\caption{
Far-field emission pattern for increasing values of the pump frequency
%$\omega_p=1.39880\,\textrm{eV}$ 
$\hbar\delta_p=-0.55\,\textrm{meV}$ 
(a),
%$1.39881\,\textrm{eV}$ 
$-0.53\,\textrm{meV}$ 
(b) below the mean-field threshold,
and 
%$1.39882\,\textrm{eV}$ 
$-0.52\,\textrm{meV}$ 
(c) close to the mean-field threshold.
The pump mode at $\kk_p$ by far saturates the grey scale, while the idler
emission is hardly visible on this scale.
The solid rectangles indicate the $\kk$-space region which forms the
signal emission. Throughout the whole paper, the correlation functions
of the selected signal emission are denoted with a bar (e.g. ${\bar
  G}\al{2}$ and ${\bar  g}\al{1,2}$).
Panel (d): cut for $k_y=0$ of the far-field emission pattern shown in the
panels (a-c). For clarity reasons, only the part of the spectrum
corresponding to the signal emission is shown.
Same cavity and pump parameters as in fig.\ref{fig:Bogo}, physical
size of the system $L_x=L_y=25\,\mu\textrm{m}$.
\label{fig:spectra_k_Below}}
\end{figure}

Below the threshold, the far-field emission patterns shown in
Fig.\ref{fig:spectra_k_Below} 
are characterized
by a strong and narrow peak corresponding to the mode at $\kk_p$ being directly
pumped by the incident laser field, and weaker and broader peaks corresponding
to the luminescence from other modes due to parametric processes
$(\kk_p,\kk_p)\longrightarrow(\kk_s,\kk_i=2\kk_p-\kk_s)$.
As the pump frequency $\omega_p$ is increased and approaches the
threshold at $\omega_p\al{c}$, not only does the intensity of the
luminescence peaks increase, but it is also spectrally narrowed in
$\kk$ space, until it becomes a very narrow peak in the vicinity of
the instability threshold. 

As discussed in Ref.\onlinecite{SqueezingTh} and experimentally verified
in Ref.\onlinecite{SqueezingExp}, the parametric luminescence has remarkable
correlations properties, in particular a significant degree of
multimode squeezing between the signal and idler.
However, as these properties do not play a central role in the
coherence properties of the signal alone, we shall not analyze them in
detail in the present paper.

\subsubsection{First- and second-order coherence}
\label{sec:Corr_below}

The real space counterpart of the linewidth reduction is an increase
of the coherence length. 
In the experiments~\cite{AugustinCoh}, a region $S$ in $\kk$-space
around the peak of the signal emission (the region inside the squares in 
fig.\ref{fig:spectra_k_Below}) is selected by means of suitable
combination of diaphragms.
The corresponding real-space correlation functions
are then defined as: 
\begin{eqnarray}
{\bar G}\al{1}(\xx,\yy)&=&\Big\langle{\hat {\bar \Psi}}_C^\dagger(\xx) \,
{\hat {\bar \Psi}}_C(\yy) \Big\rangle  \label{eq:G1} \\
{\bar G}\al{2}(\xx,\yy)&=&\Big\langle
{\hat {\bar \Psi}}_C^\dagger(\xx)\,{\hat {\bar \Psi}}_C^\dagger(\yy)\,
{\hat {\bar \Psi}}_C(\yy)\,{\hat {\bar \Psi}}_C(\xx)
\Big\rangle. \label{eq:G2}
\end{eqnarray}
in terms of the field ${\hat {\bar \Psi}}_C(\xx)$ operators
\begin{equation}
{\hat {\bar \Psi}}_C(\xx)=\sum_{\kk\in
  S}\frac{1}{\sqrt{V}}\Psih_C(\kk)\,e^{i\kk\xx}.
\end{equation}
where the sum over $\kk$ vectors is here restricted to the selected region
$S$ only. The bar over ${\bar G}\al{1,2}$ indicate that these
correlation functions are restricted to the selected signal emission only.
As usual, the translational invariance of the system guarantees that
all two point 
correlation functions depend only on the distance $\xx-\yy$.

\begin{figure}[htbp]
\includegraphics[width=3.3in]{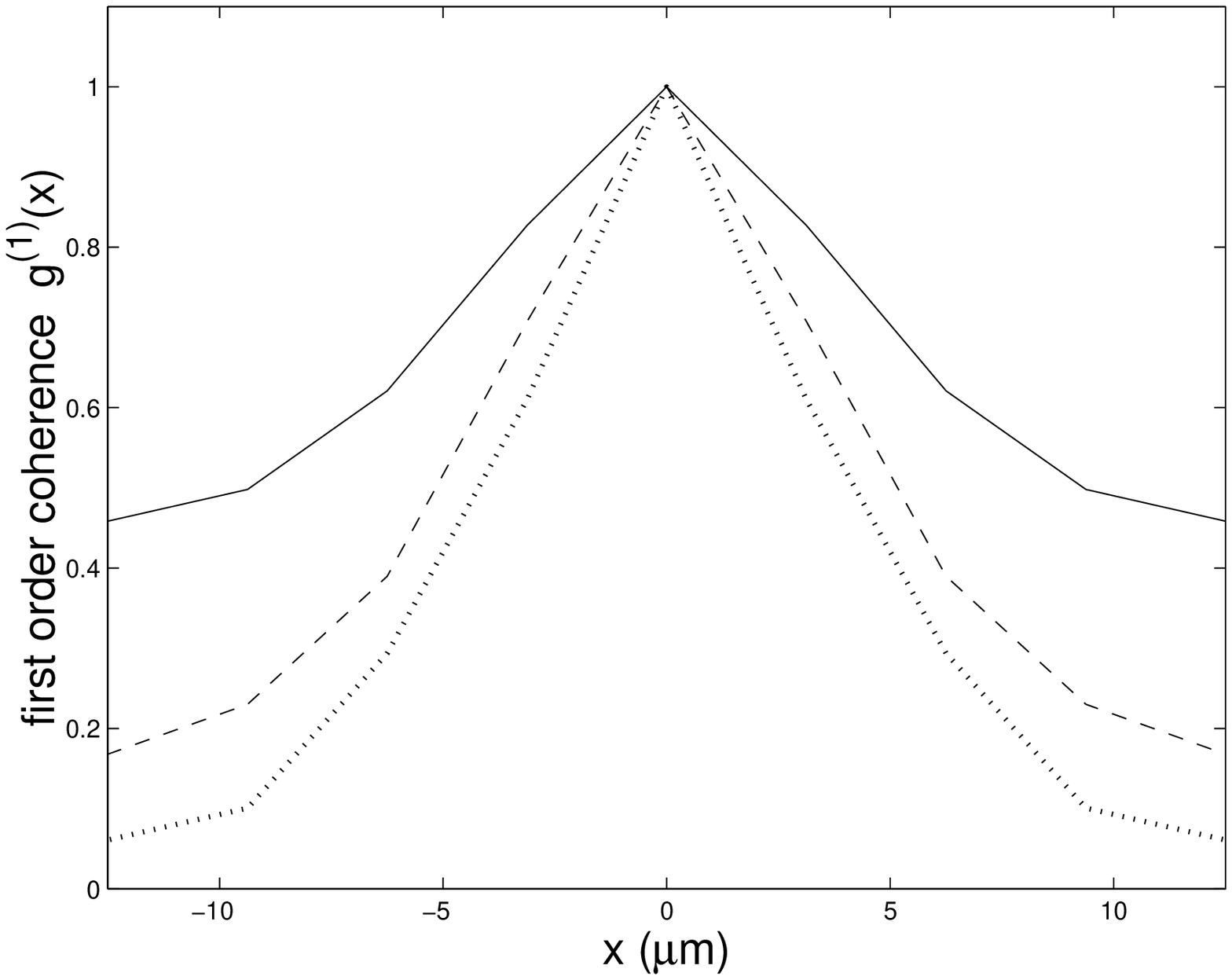}
\caption{
First-order coherence function ${\bar g}\al{1}(\xx)$ of the selected
(see solid rectangle in 
fig.\ref{fig:spectra_k_Below}) signal emission for values
of the pump frequency 
%$\omega_p=1.39880\,\textrm{eV}$
$\hbar\delta_p=-0.55\,\textrm{meV}$
(dotted), 
%$\omega_p=1.39882\,\textrm{eV}$
$\hbar\delta_p=-0.53\,\textrm{meV}$
 (dashed) and
%$1.39883\,\textrm{eV}$ 
$\hbar\delta_p=-0.52\,\textrm{meV}$
(solid)  respectively below and
around the mean-field threshold.
The cavity and the other pump parameters are the same as in
fig.\ref{fig:Bogo}.
\label{fig:g1bar_Below}}
\end{figure}

In fig.\ref{fig:g1bar_Below} we have plotted
the first order coherence function for the signal emission:
\begin{equation}
{\bar g}\al{1}(\xx)=\frac{{\bar G}\al{1}(\yy,\yy+\xx)}{{\bar
    G}\al{1}(\yy,\yy)}.
\eqname{g1bar}
\end{equation}
For $\omega_p$ well below the threshold,
${\bar g}\al{1}(\xx)$ goes from $1$ to $0$ within a transverse
coherence length $\ell_c$ which due the Wiener-Khintchine theorem is
proportional to the inverse of the $\kk$-space linewidth. 
As the threshold is approached, $\ell_c$ increases. 
At the threshold, ${\bar g}\al{1}(\xx)$ is
everywhere finite so that the coherence extends over the whole
(finite) sample.
From an experimental point of view, ${\bar g}\al{1}(\xx)$ can be
measured by first angularly filtering out the emission in the
$\kk$-space region $S$ centered around $\kk_s$,
and then performing a sort of Young's two-slit experiment like the one
in Ref.\onlinecite{AugustinCoh}.

\begin{figure}[htbp]
\includegraphics[width=3.3in]{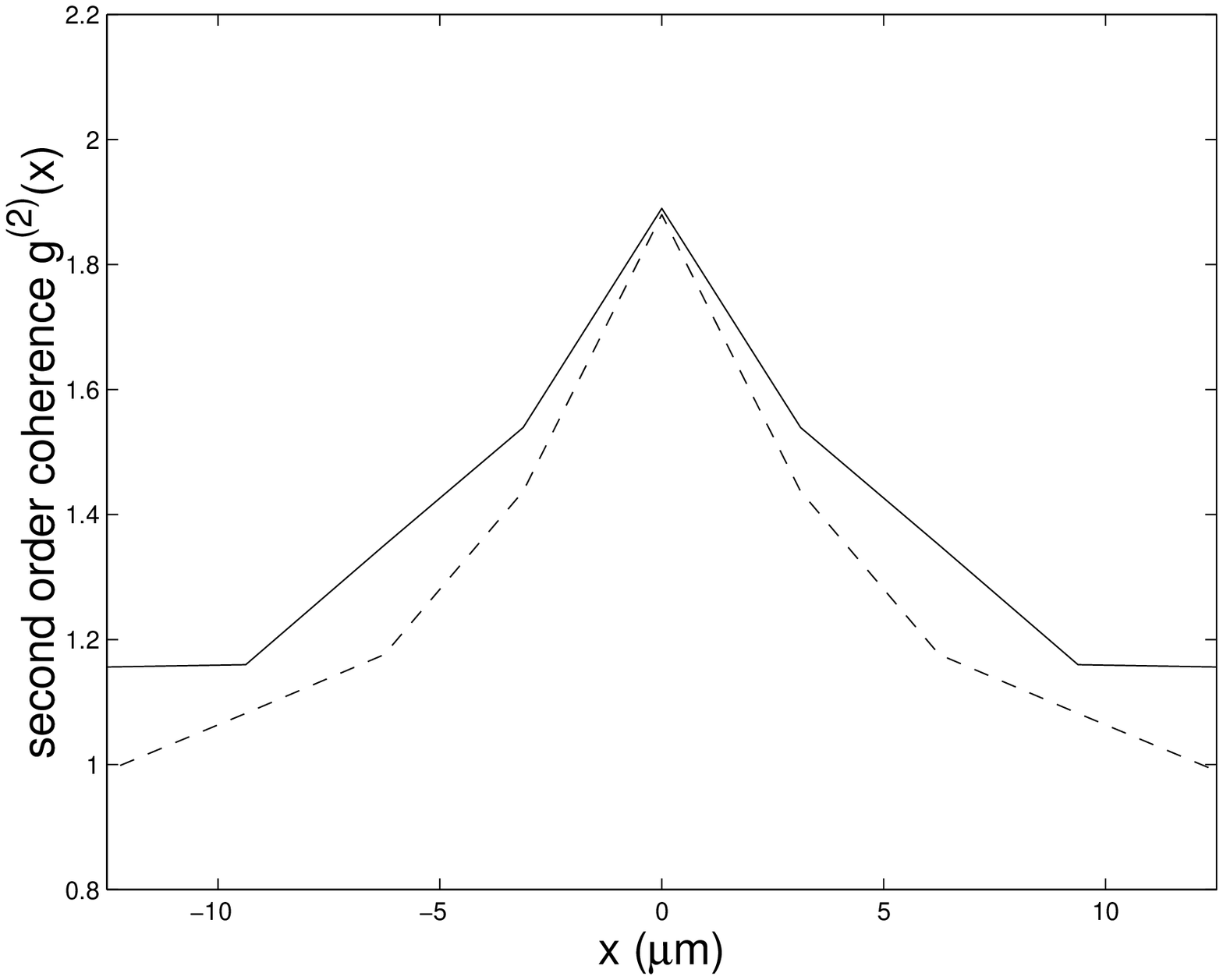}
\caption{
Second-order coherence function ${\bar g}\al{2}(\xx)$ of the selected
(see solid rectangle in 
fig.\ref{fig:spectra_k_Below}) signal emission for values of the pump
frequency  
%$\omega_p=1.39882\,\textrm{eV}$ 
$\hbar\delta_p=-0.53\,\textrm{meV}$ 
(dashed)
%$1.39883\,\textrm{eV}$ 
$-0.52\,\textrm{meV}$ 
(solid) respectively below and
around the mean-field threshold.
The cavity and the other pump parameters are the same as in
fig.\ref{fig:Bogo}.
\label{fig:g2bar_Below}}
\end{figure}

The second-order coherence function of the signal
emission ${\bar g}\al{2}$ is defined as: 
\begin{equation}
{\bar g}\al{2}(\xx)=\frac{{\bar   G}\al{2}(\yy,\yy+\xx)}{[{\bar
      G}\al{1}(\yy,\yy)]^2}
\eqname{g2bar}
\end{equation}
and experimentally can be accessed by noise correlation experiments
like the ones reported in Ref.\onlinecite{AugustinCoh}. As previously, the bar over
${\bar g}\al{2}$ indicates that this coherence function includes
the selected signal emission only.

For $\omega_p$ well below the threshold $\omega_p\al{c}$, ${\bar
  g}\al{2}(\xx)$ 
  shows the typical Hanbury-Brown and Twiss bunching~\cite{HB-T} at
  short distances $|\xx|\ll \ell_c$, where its value is close to
  $2$, and then it tends to $1$ for $|\xx|\gg \ell_c$. 
This behaviour can be easily understood in terms of the Gaussian
structure of the field: well below the threshold, all modes of the coupled
excitonic and photonic fields in the microcavity are in a Gaussian
  state, exception made for the one at $\kk_p$ directly driven by the
  pump. 
This means that Wick theorem applies and higher-order correlation
  functions can be consequently factorized into products of
  first-order ones~\cite{LowD}. 
%, e.g.
%\begin{multline}
%\big\langle \Psihd_C(\xx)\,\Psihd_C(\xx')\,
%\Psihd_C(\xx')\,\Psih_C(\xx) \big\rangle
%=\\
%=\big\langle\Psihd_C(\xx)\,\Psih_C(\xx) \big\rangle
%\big\langle\Psihd_C(\xx')\,\Psih_C(\xx')  \big\rangle
%+\\
%+\big\langle \Psihd_C(\xx')\,\Psih_C(\xx) \big\rangle
%\big\langle \Psihd_C(\xx)\,\Psih_C(\xx') \big\rangle
%+\\
%+\big\langle \Psihd_C(\xx)\,\Psihd_C(\xx') \big\rangle
%\big\langle \Psih_C(\xx')\,\Psih_C(\xx) \big\rangle.
%\end{multline}
As the parametric emission gives
non-vanishing anomalous correlations only between the signal and the
idler modes, all the anomalous correlations appearing in the
Wick factorization of the signal second-order correlation function
${\bar G}\al{2}(\xx,\yy)$ vanish and this can be factorized as: 
\begin{equation}
{\bar G}\al{2}(\xx,\yy)\simeq{\bar G}\al{1}(\xx,\xx)\,{\bar
  G}\al{1}(\yy,\yy)+
\big|{\bar G}\al{1}(\xx,\yy)\big|^2
\eqname{G2bar}
\end{equation}
so that ${\bar g}\al{2}(\xx)$ is
\begin{equation}
{\bar g}\al{2}(\xx)\simeq
1+\big| {\bar g}\al{1}(\xx)\big|^2.
\eqname{g2bar_decomp}
\end{equation}
This form is consistent with the numerical results
shown in Figs.\ref{fig:g1bar_Below} and \ref{fig:g2bar_Below}. In
particular, the coherence length of ${\bar G}\al{2}$ is roughly half
the one of ${\bar G}\al{1}$. Note that the decomposition \eq{G2bar} would
not be valid for quantities without a bar, e.g. $G\al{2}(\xx)$ that
will be introduced in \eq{g2x}) since in that case the anomalous
correlation between signal and idler gives an important contribution.

Closer to the threshold, the field is no longer in a Gaussian state:
  the short-distance value ${\bar  g}\al{2}(0)$ which describes the
  intensity fluctuations of the signal emission is reduced to a value
  close to $1$ as it   happens for a coherent field.

\subsection{Above the parametric instability threshold}

\subsubsection{Far-field angular pattern}

\begin{figure}[htbp]
\includegraphics[width=3.3in]{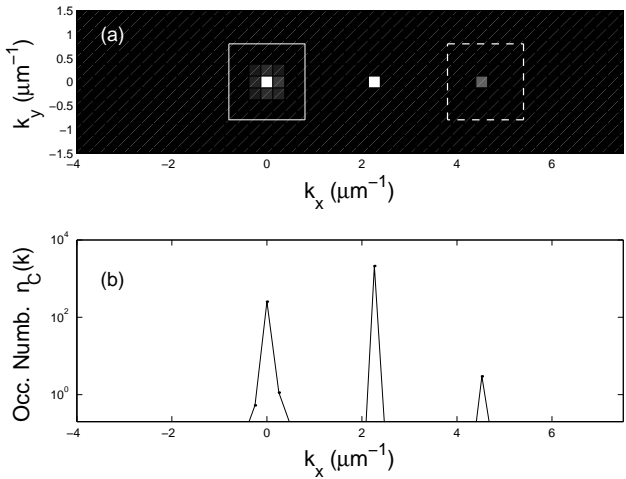}
\caption{
Upper (a) panel: far-field emission pattern for a pump frequency
%$\omega_p=1.39885\,\textrm{eV}$  
$\hbar\delta_p=-0.5\,\textrm{meV}$  
(above the mean-field threshold).
The solid (dashed) rectangles indicate the $\kk$-space region which forms the
signal (idler) emission. Throughout the whole paper, the correlation functions
of the selected signal emission are denoted with a bar (e.g. ${\bar
  G}\al{2}$ and ${\bar 
  g}\al{1,2}$). 
Lower (b) panel: cut for $k_y=0$ of the far-field emission pattern
shown in panel (a).
The cavity and the other pump parameters are the same as in
fig.\ref{fig:Bogo}.
\label{fig:spectra_k_Above}}
\end{figure}

As the pump frequency $\omega_p$ goes through the critical value 
$\omega_p\al{c}$, the signal and idler emission concentrates in a
single signal/idler pair of modes at $\kk_{s,i}$ and their intensity
becomes of the same order as the one in the pump mode at $\kk_p$
(Fig.\ref{fig:spectra_k_Above}).

\begin{figure}[htbp]
\includegraphics[width=3.3in]{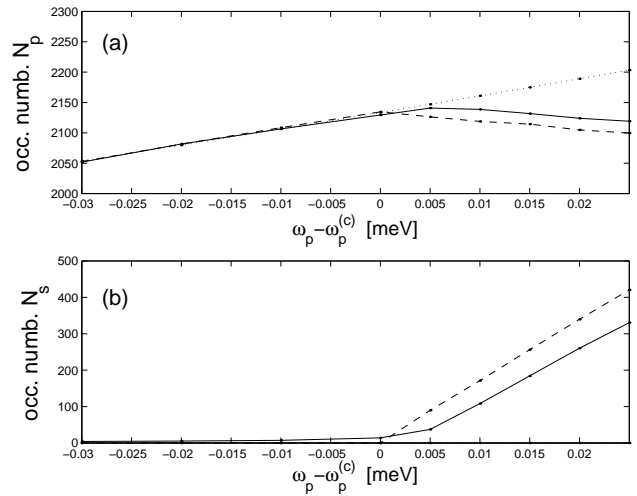}
\caption{
Intensity in the pump (a) and signal (b) modes as a function of the
pump frequency $\omega_p$. Solid lines: Wigner-Monte Carlo results.
Dashed lines: mean-field prediction. The dotted line in panel (a)
gives the intensity in the pump mode for the unstable homogeneous
mean-field state of (\ref{eq:ss_X}-\ref{eq:ss_C}).
The cavity and the other pump parameters are the same as in
fig.\ref{fig:Bogo}.
\label{fig:threshold}}
\end{figure}

In Fig.\ref{fig:threshold} we have plotted the intensity in the pump
and the signal modes as a function of the pump frequency $\omega_p$. 
While the pump intensity $N_p$ is simply the occupation number $n_C(\kk_p)$
of the mode at $\kk_p$, the signal and idler intensities $N_{s,i}$ are defined as
the sum of the occupation numbers $n_C(\kk)$ for $\kk$ within the
regions contained in respectively the solid and dashed rectangles in
Fig.\ref{fig:spectra_k_Above}.  

Both the pump and the signal intensities show a threshold behaviour at
$\omega_p\al{c}$: the pump intensity $N_p$ monotonically increases
upto the threshold, and then starts slightly decreasing because of the
presence of the finite signal [fig.\ref{fig:threshold}(a)]. 
The signal one $N_s$ is very weak below the threshold, and has a
sudden increase close to $\omega_p\al{c}$[fig.\ref{fig:threshold}(b)]. 
As compared to the
mean-field prediction~\footnote{This can be numerically obtained by means of the same
Wigner-Monte Carlo code by taking the mean-field
limit~\cite{CastinHouches99} $g\rightarrow 
0$, $E_p\rightarrow \infty$ at a fixed $g\,|E_p|^2$ (i.e. at a fixed
interaction energy $g\,|\psi_X|^2$).}, 
the transition is rounded by quantum fluctuations, which give indeed a finite signal
intensity also below the threshold and smoothen the threshold. 
Note that fluctuations are also responsible for the slight shift of
the threshold which can be observed in the figure. The behaviour of
the idler intensity (not shown) is strictly analogous to the one of
the signal one.

Far above the threshold, additional, weaker, peaks appear due to multiple 
scattering process~\cite{Bragg}, but they are filtered out when the signal
emission is selected.
A discussion of these coherent peaks, as well as of the novel spectral and
angular features shown by the weak incoherent luminescence when we are
far above the threshold go beyond the scope of the present paper and
will be postponed to future work.

\subsubsection{First- and second-order coherence function}

As in sec.\ref{sec:Corr_below}, we have selected the region $S$ in
wavevector space corresponding to the signal emission (within the
solid rectanble in Fig.\ref{fig:spectra_k_Above}) and we have
determined its coherence properties.

\begin{figure}[htbp]
\includegraphics[width=3.3in]{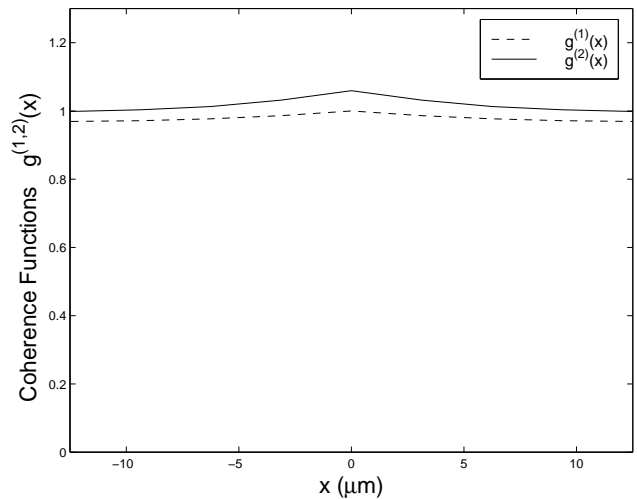}
\caption{
First- (dashed) and second-order (solid) coherence functions 
${\bar g}\al{1,2}(\xx)$ of the
selected (see solid rectangle in
fig.\ref{fig:spectra_k_Below}) signal emission for a pump frequency
%$\omega_p\al{c}=1.39885$ 
$\hbar\delta_p\al{c}=-0.5\,\textrm{meV}$ 
(above the mean-field threshold). 
The cavity and the other pump parameters are the same as in
fig.\ref{fig:Bogo}.
\label{fig:g12bar_Above}}
\end{figure}

The first- and
second-order coherence functions ${\bar g}\al{1,2}(\xx)$ are plotted
in fig.\ref{fig:g12bar_Above}.
${\bar g}\al{1}(\xx)$ is nearly flat and equal to $1$, which indicates
that the first-order coherence extends over the whole two-dimensional
sample, or, in other terms, that the emission is phase coherent.
As expected for a coherent field, the second-order coherence function
  ${\bar g}\al{2}(\xx)$ is also almost flat and its value close to
  $1$: in physical terms, this means that the intensity fluctuations
of the signal emission are weak. 

Both these qualitative behaviours are in agreement with the
experimental observations reported in Ref.\onlinecite{AugustinCoh}.

\section{Analogy with the Bose gas at thermal equilibrium}
\label{sec:AnalogyBEC}

The analogy between the threshold of a optical parametric oscillator
or of a laser and a sort of non-equilibrium second-order phase transition
was put forward in the '70s~\cite{OptPhaseTrans,ScullyZubairy},
but its study was mostly focussed on the case of few-mode systems,
such as optical resonators with discrete modes.
Its generalization to spatially extended systems showing a continuum
of modes is instead a much more recent topic from both the
theoretical~\cite{QuantumImagesOPO} and the
experimental~\cite{AugustinCoh} points of view.
In the present section, we shall try to interpret the results
presented in the previous section in terms of a sort of
non-equilibrium Bose-Einstein condensation.

At the level of mean-field theory, the analogy is immediate: in usual
statistical mechanics, the onset of a phase transition is signalled
by the thermodynamical instability of the symmetric state and the
consequent appearance of a non-vanishing value of the order parameter
in the equilibrium state.
For the specific case of the Bose-Einstein condensation transition,
the order parameter consists of the expectation value of the Bose
field $\Psih(\xx)$.

In the present non-equilibrium case, the mean-field threshold
corresponds to the point where the homogeneous solution
\eq{ss_X} and \eq{ss_C} becomes unstable and a finite signal and idler
amplitudes appear in the new stable stationary state.
As in the standard theory of the optical parametric oscillator, the
absolute phases of the signal and the idler are not determined. The
Hamiltonian is in fact invariant under the $U(1)$ symmetry which sends
$\psi\al{s,i}\rightarrow\psi\al{s,i}\,e^{\pm i\theta}$ and the specific
value of the phase is randomly chosen at each realization by the
spontaneous symmetry breaking mechanism. The chosen value of the
order parameter is the same throughout the whole system. 

Quantitative results to illustrate the physics of this spontaneous
symmetry breaking are obtained from the Wigner-Monte Carlo
calculations including the fluctuations around the mean-field which
smoothen the threshold.  
Because of the fluctuations, the signal and idler amplitudes no longer
vanish below threshold, but as the signal and idler emissions take
place into many available modes which are close to resonance, they are
incoherent both spatially and temporally. 
As the  threshold is approached, the characteristic coherence length
which determines the spatial behaviour of the first- and second-order
coherence functions ${\bar g}\al{1,2}(\xx)$ increases and becomes
macroscopic at the threshold. Simultaneously, the intensity
fluctuations, quantified by ${\bar   g}\al{2}(0)$, are quenched. 

Above the threshold, the parametric emission is concentrated in a single
pair of modes, whose amplitude is accurately described by a
${\mathbf C}$-number as in the mean-field theory. Fluctuations give
only a small correction.
In particular, the signal and idler phases are constant throughout the
whole two-dimensional sample. As it is usual in the spontaneous
symmetry breaking phenomena, this does not violate the $U(1)$ symmetry
of the system Hamiltonian, since the phase is chosen at random.

These behaviours are strictly analogous to what happens in a
Bose gas at thermal equilibrium when the temperature is lowered across
the Bose-Einstein condensation  temperature $T_{BEC}$. 
For $T>T_{BEC}$, the gas is a thermal, incoherent, one and is spread
among many low-lying modes.
Its coherence extends for a finite distance, generally smaller than
the interparticle spacing, and the same coherence length
determines the spatial behaviour of both the phase (first-) and
the density-density (second-order) correlation function of the gas.
For $T\rightarrow T_{BEC}^+$, the coherence length
diverges~\cite{AtomicBEC,CohBose}.
For $T<T_{BEC}$ a macroscopic population is concentrated in
the single state of lowest energy and an off-diagonal long-range order
appears in the real-space one-body density matrix. This is the so-called
Penrose-Onsager criterion for Bose-Einstein
condensation~\cite{PenroseOnsager}.

Also from the experimental point of view there is a close relation
between experiments like the ones in Ref.\onlinecite{AugustinCoh} which measure
the coherence of the light emission, and experiments which address the
coherence of the atomic matter waves. 
The first-order coherence of
the matter wave of an atomic BEC has in fact been revealed by looking at
the interference fringes formed by two atomic wavepackets extracted at 
different spatial positions~\cite{Bloch_g1}, while the density
fluctuations (quantified by $g\al{2,3}(0)$) have been 
measured from the interaction energy~\cite{2body} and the three-body
recombination rate~\cite{3body}: both of them are quenched 
as $T$ is lowered below $T_{BEC}$.

Although the analogy is very close, it is important to stress the
fundamental differences between the Bose-Einstein condensation phenomenon of
equilibrium statistical mechanics and the appearance of spontaneous
coherence in the parametric emission from the present microcavity system.
As polaritons have a finite lifetime, they have to be continuously
injected in the system from an external pump, and the stationary state
arises as a sort of dynamical equilibrium between the injection and
the dissipation. Instead of a true thermodynamical equilibrium, we are
therefore dealing with the stationary state of a {\em
  driven-dissipative} system~\cite{PatternFormation}.
Hence, there is no reason for the critical properties to be the same, and the
general theorems of equilibrium statistical mechanics do not
necessarily apply.

\section{Modulational instability and quantum images}
\label{sec:Modulational}

An alternative interpretation of the parametric instability can be put
forward in terms of modulational instability, the initially spatially 
homogeneous intensity profile becoming unstable with respect to a
periodic modulation which breaks the translational symmetry. 
Behaviours of this kind are well-known in the
framework of nonlinear dynamics and pattern
formation~\cite{PatternFormation}, and have analogies with the
transition from the liquid to a cristalline solid state.
The concept of modulational instability has recently 
played a central role in the understanding 
of experiments with atomic Bose-Einstein
condensates~\cite{ModInstBEC}.

The physical origin of the intensity modulation and
its relation with the parametric instability can be understood by
looking at the intensity modulation pattern formed by the interference
of the pump, the signal and the idler. 
In $\kk$ space, the pump, the signal and the idler are localized
around respectively $\kk_p$, $\kk_s$ and $\kk_i=2\kk_p-\kk_s$ and the
polariton wavefunction can be written as: 
\begin{equation}
\psi_i(\xx)=\psi\al{p}_i\,e^{i\kk_p\xx}+
\psi\al{s}_i(\xx)\,e^{i\kk_s\xx}+
\psi_i\al{i}(\xx)\,e^{i\kk_i\xx},
\eqname{psi_splitting}
\end{equation}
where $\psi_i\al{s,i}(\xx)$ are slowly varying functions as compared
to the period $\ell_\rho=2\pi/|\kk_p-\kk_s|$ of the intensity modulation.
As a shift of the signal and idler phases by $\pm\phi$ corresponds to
a spatial translation of the intensity modulation pattern by $\Delta
x=\ell_\rho \phi / 2\pi$ along the direction of $\kk_s-\kk_p$, the
spontaneous breaking of the $U(1)$ symmetry corresponding to the
signal and idler phases can also be interpreted as the spontaneous
breaking of the translational symmetry along the direction of $\kk_p-\kk_s$.
As the position of the fringes is random, the mean intensity
profile remains flat, which guarantees that the translational
invariance of the system Hamiltonian \eq{Hamilt_tot} is preserved.

\begin{figure}[htbp]
\includegraphics[width=2.5in]{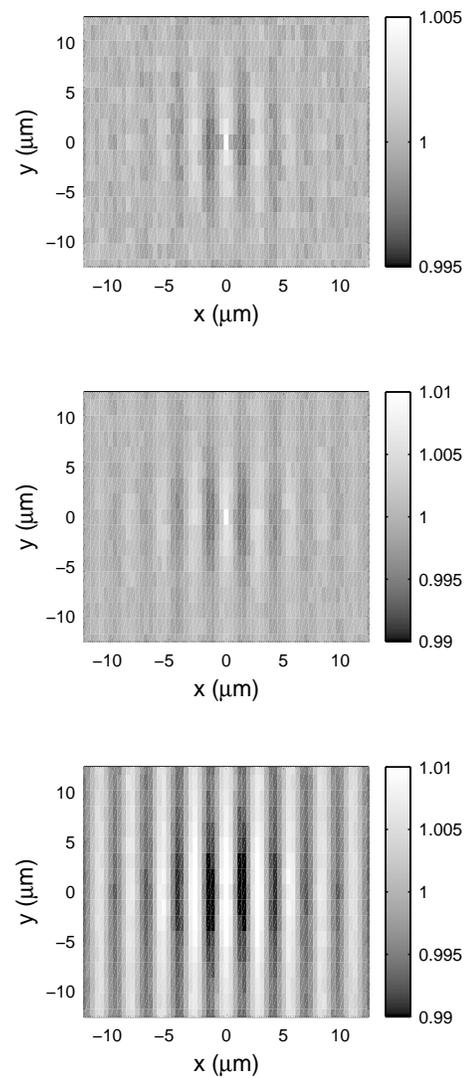}
\caption{
Intensity correlation function $G\al{2}(\xx)$ of the in-cavity
polariton field for values of the pump frequency
%$\omega_p=1.39880\,\textrm{eV}$ 
$\hbar\delta_p=-0.55\,\textrm{meV}$ 
(upper panel),
%$\omega_p=1.39882\,\textrm{eV}$ 
$\hbar\delta_p=-0.53\,\textrm{meV}$ 
(central panel) 
and
%$\omega_p=1.39883\,\textrm{eV}$
$\hbar\delta_p=-0.52\,\textrm{meV}$  
(lower panel) respectively below and
around the mean-field threshold.
The cavity and the other pump parameters are the same as in
fig.\ref{fig:Bogo}.
\label{fig:g2x_Below}}
\end{figure}

\begin{figure}[htbp]
\includegraphics[width=3.in]{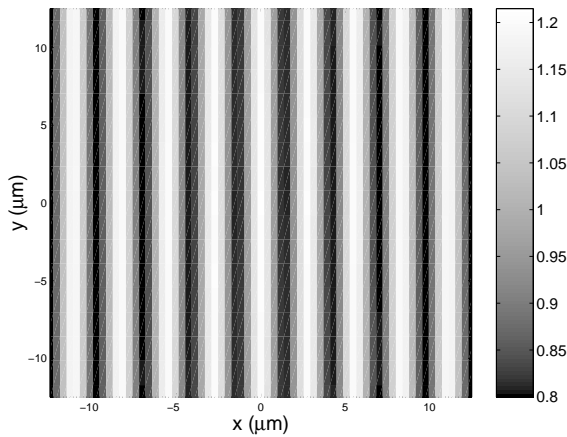}
\caption{
Intensity correlation function $G\al{2}(\xx)$ of the in-cavity polariton field
for a pump frequency 
%$\omega_p=1.39885\,\textrm{eV}$ 
$\hbar\delta_p=-0.5\,\textrm{meV}$ 
above the threshold. 
The cavity and the other pump parameters are the same as in
fig.\ref{fig:Bogo}.
\label{fig:g2x_Above}}
\end{figure}

On the other hand, fringes appear clearly in the intensity correlation
function of the polariton field
\begin{equation}
  \label{eq:g2x}
  G\al{2}(\xx)=
\Big\langle\Psihd_C(\yy)\,\Psihd_C(\yy+\xx)\,
\Psih_C(\yy+\xx)\,\Psih_C(\yy) \Big\rangle,
\end{equation}
which is plotted in fig.\ref{fig:g2x_Below}. 
Experimentally, $G\al{2}(\xx)$ can be measured by simply isolating
in the near-field the light emitted from a pair of points $\yy$ and
$\yy+\xx$ spaced of
$\xx$ and then correlating the corresponding intensity fluctuations.
As the intensity pattern arises from the interference of signal, pump and
idler modes, no preliminar angular selection must now be performed and,
for this reason, the symbol $G\al{2}(\xx)$ does not carry a bar over
it.  

A detailed study of the appearance of the fringes, effect which in
quantum optics goes under the name of {\em quantum
  images}~\cite{QuantumImages}, was given in Ref.\onlinecite{QuantumImagesOPO}
for the simple case of an optical 
parametric oscillator formed by a $\chi\al{2}$ medium in a planar
cavity. 
Here we want to emphasize the relation between the phase coherence
properties of the signal/idler emission, quantified by ${\bar
  g}\al{1}(\xx)$ defined in \eq{g1bar} and the coherence length of the
periodic modulation of $G\al{2}(\xx)$.
As a shift  of the signal and idler phases by
$\pm\phi$ corresponds to a spatial translation displacement of the intensity
pattern, a spatial rigidity of the phase of $\psi\al{s,i}(\xx)$ transfers into
a corresponding rigidity of the intensity modulation pattern. On the
other hand, when the phase of $\psi\al{s,i}(\xx)$ is able to freely
fluctuate in space, the intensity modulation pattern is washed out.
This can be observed in the Figs.\ref{fig:g2x_Below} and
\ref{fig:g2x_Above} which show $G\al{2}(\xx)$ for the two cases 
respectively below and above the threshold: in the first case, the
spatial extension of the modulation pattern is finite and equal to the
phase coherence length $\ell_c$, while in the second case  the pattern
extends over the whole system.
A related interference effect has been used in Ref.\onlinecite{quasicond2} to
infer the coherence length of an atomic Bose-Einstein condensate from a
measurement of intensity correlations.

\begin{figure}[htbp]
\includegraphics[width=2.5in]{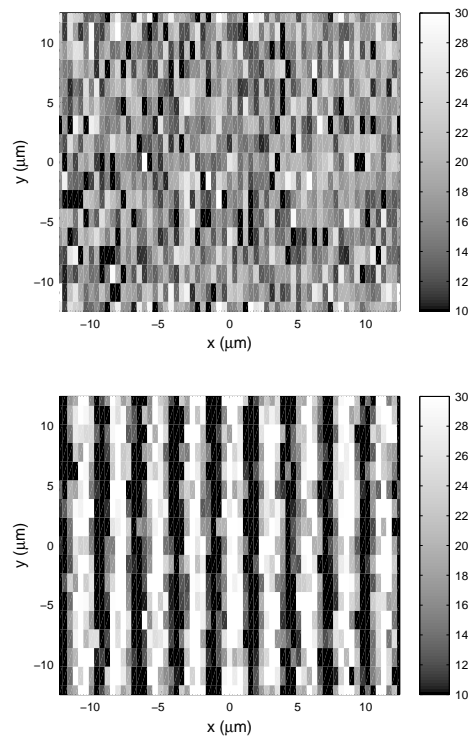}
\caption{
Intensity profile $|\psi_C(\xx)|^2|$ of a single realization for pump
frequencies 
%$\omega_p=1.39882\,\textrm{eV}$ 
$\hbar\delta_p=-0.53\,\textrm{meV}$ 
(upper panel) and
%$\omega_p=1.39885\,\textrm{eV}$ 
$\hbar\delta_p=-0.5\,\textrm{meV}$ 
(lower panel) respectively below and above the mean-field threshold.
The cavity and the other pump parameters are the same as in
fig.\ref{fig:Bogo}, exception made for the coupling constant $g$ and
the driving 
intensity $I_p$ respectively increased and decreased by a factor 
$5$ so as to clarify the picture without affecting the mean-field physics.
\label{fig:e_x_single}}
\end{figure}

In Ref.\onlinecite{ZambriniPRA}, the spontaneous breaking of translational symmetry in a
planar optical parametric oscillator was pointed out as a slowing down
of the diffusive motion of the modulation pattern during the time
evolution of a single realization. 
Here we show another interesting consequence of the same phenomenon
in the intensity profile of a single realization of the Wigner-Monte
Carlo, that is in an instantanous measurement of the local intensity.
While below threshold the modulation pattern is too weak to be
detected on a single realisation and has to be obtained by taking the 
average over many realisations, it appears clearly although somewhat noisy in
a picture taken above threshold when the occupation of the signal and
idler modes is large and coherent
(fig.\ref{fig:e_x_single}). Obviously, the position of the
pattern is random and varies from one realisation to the next.

\section{One-dimensional system: local vs. global coherence}
\label{sec:1D}

All the physical discussions presented in the previous sections were based on a 
mean-field picture: below threshold, many modes are incoherently
populated, while above threshold, there is a macroscopic occupation of
a single signal/idler pair of mode and fluctuations give only a small correction.
This interpretation was validated by the numerical results of the
complete Wigner-Monte Carlo simulations which were performed for
two-dimensional systems of side $L_x=L_y=25\,\mu\textrm{m}$: above the
threshold, coherence of the signal emission extended over the whole
two-dimensional sample giving a sort of non-equilibrium long-range order.

\begin{figure}[htbp]
\includegraphics[width=3.3in]{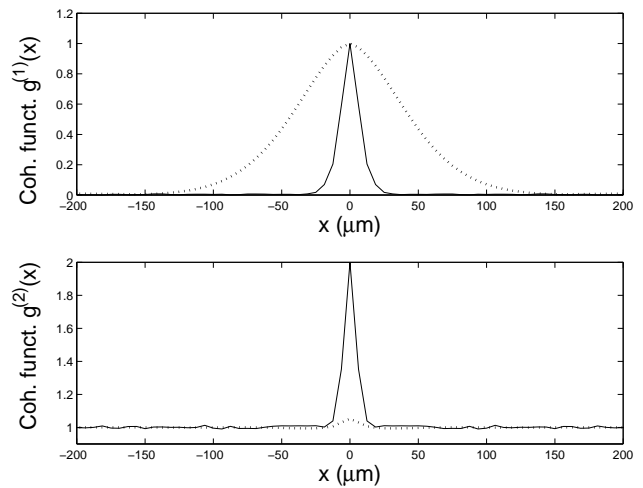}
\caption{
First- (upper panel) and second-order (lower panel) coherence
functions ${\bar g}\al{1,2}(\xx)$ of the
selected signal emission for pump frequencies respectively
%$\omega_p\al{c}=1.39882$ 
$\hbar\delta_p=-0.53\,\textrm{meV}$
below the mean-field threshold (solid) and 
%$\omega_p\al{c}=1.39884$
$\hbar\delta_p=-0.51\,\textrm{meV}$
above the mean-field threshold (dotted).
The cavity and the other pump parameters are the same as in
fig.\ref{fig:Bogo}, 
exception made for the geometry: the system is here one-dimensional
and $L_x=400\,\mu\textrm{m}$.
\label{fig:quasiBEC}}
\end{figure}

It is a well-known fact of equilibrium statistical mechanics that the
spontaneous breaking of a $U(1)$ symmetry and the appearance of a
long-range order is possible only in dimension $d>2$ at finite
temperature and in $d\geq 2$ at zero temperature (Mermin-Wagner
theorem~\cite{Huang,MerminWagner}), while otherwise coherence extends
over a finite distance only.
In order to investigate the possibility of observing such behaviours in a
non-equilibrium context, we have performed Wigner-Monte Carlo
simulations for a system with the same pump parameters as in the previous
sections,  but in a completely different cavity geometry. 
Instead of a square of size $L_x=L_y=25\,\mu\textrm{m}$, we
consider here a one-dimensional system of length $L_x=400\,\mu\textrm{m}$.
Along this axial direction, the polaritons are free and  periodic boundary
conditions are assumed, while the transverse motion is assumed to be
frozen in the lowest state by a strong lateral confinement.
This kind of reduced-dimensionality systems are well within the
possibilities of actual semiconductor technology: 
effectively one-dimensional systems has been recently demonstrated by
several groups~\cite{PhotonWires}. 

The results of numerical calculations for the first- and second-order
coherence of the signal emission are shown in fig.\ref{fig:quasiBEC}:
below the threshold, the behaviour is similar to the one already
observed in the two-dimensional case (figs.\ref{fig:g1bar_Below} and
\ref{fig:g2bar_Below}): the phase coherence extends over a finite
distance and intensity fluctuations due to the Hanbury-Brown and Twiss
bunching effect are important ${\bar g}\al{2}\approx 2$.

Above the threshold, the intensity fluctuations are strongly
suppressed ${\bar g}\al{2}\approx 1$ as in the two-dimensional case of
fig.\ref{fig:g12bar_Above}, while the phase coherence of the signal
extends over a longer but still finite distance. The large 
distance limit of ${\bar g}\al{1}$ is in fact $0$ and no long-range
coherence exists. 
Similar results were discussed in \cite{ZambriniPRA} for 
the case of a roll-pattern instability in an optical cavity.

In the language of equilibrium statistical mechanics, phenomena of
this kind are well-known and usually go under the name of {\em
  quasi-condensates}~\cite{LowD,quasicond,quasicond2}, so to stress the fact that a local 
order is present (signalled by the reduced value of the intensity
fluctuations), but not the global one which characterizes a true Bose-Einstein
condensate. The long range order is in fact destroyed by the long
wavelength fluctuations of the order parameter, in our case the local phase
of the signal emission
\footnote{In principle, a similar quasi-condensation behaviour may show up in
two-dimensional systems as well, but the distinction between a
condensate and a quasi-condensate may require very exponentially large
samples to rule out the possibility of a finite-size-induced
BEC~\cite{AtomicBEC}. Analytical work in this direction is under way.
}.

A final remark: recently, the realization of a BEC of polaritons was claimed
in Ref.\onlinecite{Yama_g2} starting from the observation of a reduced value of
${\bar g}\al{2}$. 
Unless a detailed analysis of the geometry of the system is
performed, this observation does not appear as being a conclusive proof
of polariton BEC, as the mere reduction of intensity fluctuations does
not univocally correspond to the presence of a long-range order.
In other terms, the possibility of having a quasi-condensate of
polaritons has not been ruled out by the available experimental data.

\section{Conclusions and perspectives}
\label{sec:Conclu}

In the present paper we have presented a comprehensive study of the
coherence properties of the parametric emission from a planar
semiconductor microcavity for pump parameters spanning across the
threshold for the parametric oscillation. 

The numerical calculations have been performed by means of a Monte Carlo
technique based on the stochastic differential equations of the Wigner
representation of the coupled exciton and cavity-photon quantum fields.
Inspired by the first experimental results reported
in Ref.\onlinecite{AugustinCoh}, we have calculated the coherence properties of
the in-cavity polaritonic field and, more specifically, of the light
which is then emitted in the signal beam. In particular, we have 
analyzed the behaviour of the first and the second-order coherence
functions when the pump parameters cross the threshold for the
parametric oscillation.
The numerical results are physically understood in terms of a
non-equilibrium phase transition occurring at the parametric
threshold, a phase transition which closely ressembles to a
non-equilibrium Bose-Einstein condensation.  

Below threshold, the emission is an incoherent one, and the first- and
second-order correlation functions show a finite coherence length
for both the phase correlation and the intensity fluctuations.
As the pump approaches the threshold, the coherence length becomes
larger and larger and eventually becomes macroscopic at the threshold.
In momentum space, this corresponds to a broad momentum distribution
which gets narrowed into a delta-like peak at the threshold.

Above threshold, the emission is instead a coherent one, both the
signal and the idler intensities becoming of the same order as the
pump one, and the corresponding momentum distribution being sharply
peaked.
For two-dimensional systems of size comparable to the one of actual
experiments with planar microcavities, first- and second-order
coherence extends in the whole system.
For systems of reduced dimensionality such as photon wires, the
numerical results suggest that a local order persists in the sense
that intensity fluctuations are 
strongly reduced, but no true long-range order is present, as the global
phase coherence is destroyed by long-wavelength phase fluctuations.
This phenomenon is a non-equilibrium analog of the so-called
quasi-condensation phenomenon of reduced-dimensionality Bose gases at
equilibrium.

Although the results of the present paper have been obtained for a
very specific system of actual experimental interest, we expect the
main properties of the critical point at the threshold to be very
general for a whole class of non-equilibrium systems.
Also the absence of long-range order in low-dimensional systems is
expected to be a general fact in non-equilibrium systems:
inspired by the Mermin-Wagner theorem of equilibrium statistical
mechanics, we are actually trying to characterize the possibility of a
long-range order in non-equilibrium systems of reduced dimensionality,
and to determine the corresponding phase coherence length.
In order to have a closer comparison with experiments, the effect of a
disorder (which can be straightforwardly taken into account via the
potentials $V_{X,C}(\xx)$ in Eq.\ref{eq:Hamilt_tot}) on the
non-equilibrium critical properties will be adressed 
as well.

\begin{acknowledgments}
Continuous stimulating discussions with A. Baas, O. El Da\"\i f,
J. Tignon, G. Dasbach, C. Diederichs, A. Verger, Y. Castin, C. Lobo,
A. Sinatra, C. Tozzo, F. Dalfovo, V. Savona, and A. Rosso are warmly
acknowledged.  
I.C. is grateful to M. Wouters for his active interest on polaritonic
quasi-condensation issues.

\end{acknowledgments}

%\appendix

%\section{}

%\appendix

%\section{The third-order derivative terms in the pseudo-Fokker-Planck
%  equation for the Wigner distribution}

%\label{sec:Appendix}

\end{document}